\documentclass[11pt]{article}

\usepackage{url}
\usepackage{color}
\usepackage{graphicx}
\usepackage{bm}
\usepackage{amsmath}
\usepackage{amssymb}
\DeclareMathOperator{\Tr}{Tr}

\usepackage{setspace}
\newenvironment{sabstract}{%
\begin{quote} }
{\end{quote}}

\title{\vspace*{-2cm} A change in stripes for cholesteric shells via anchoring in moderation}

\author
{Lisa Tran,$^{1}$ Maxim O. Lavrentovich,$^{1}$ Guillaume Durey,$^{2}$ \\
Alexandre Darmon,$^{2}$ Martin F. Haase,$^{3}$ Ningwei Li,$^{4}$ \\
Daeyeon Lee,$^{4}$ Kathleen J. Stebe,$^{4}$ \\
Randall D. Kamien,$^{1\ast}$ and Teresa Lopez-Leon$^{2\ast}$ \\
\\
\\
\small{$^{1}$Department of Physics and Astronomy, University of Pennsylvania,}\\
\small{Philadelphia, PA 19104, USA}\\
\small{$^{2}$UMR CNRS 7083, \'{E}cole sup\'{e}rieure de physique et de chimie industrielles de la Ville de Paris,}\\
\small{PSL Research University, 75005 Paris, FR}\\
\small{$^{3}$Department of Chemical Engineering, Rowan University,}\\
\small{Glassboro, NJ 08028, USA}\\
\small{$^{4}$Department of Chemical and Biomolecular Engineering, University of Pennsylvania,}\\
\small{Philadelphia, PA 19104, USA}\\
\\
\normalsize{$^\ast$To whom correspondence should be addressed;}\\
\normalsize{E-mail: kamien@upenn.edu, teresa.lopez-leon@espci.fr.}
}

\date{}

\begin{document} 

% Double-space the manuscript.

%\baselineskip24pt

% Make the title.
\maketitle

% Place your abstract within the special {sciabstract} environment.

\begin{sabstract}
Chirality, ubiquitous in complex biological systems, can be controlled and quantified in synthetic materials such as cholesteric liquid crystal (CLC) systems. In this work, we study spherical shells of CLC under weak anchoring conditions. We induce anchoring transitions at the inner and outer boundaries using two independent methods: by changing the surfactant concentration or by raising the temperature close to the clearing point. The shell confinement leads to new states and associated surface structures: a state where large stripes on the shell can be filled with smaller, perpendicular sub-stripes, and a focal conic domain (FCD) state, where thin stripes wrap into at least two, topologically required, double spirals. Focusing on the latter state, we use a Landau-de Gennes model of the CLC to simulate its detailed configurations as a function of anchoring strength. By abruptly changing the topological constraints on the shell, we are able to study the interconversion between director defects and pitch defects, a phenomenon usually restricted by the complexity of the cholesteric phase. This work extends the knowledge of cholesteric patterns, structures that not only have potential for use as intricate, self-assembly blueprints but are pervasive in biological systems.
\end{sabstract}

\newpage

Chiral biological building blocks, such as DNA \cite{livolant-dna, clark}, viruses \cite{zvonfrad}, and chitin \cite{chitin}, often form self-organised helical structures, called cholesteric liquid crystal (CLC) phases. In these phases, the orientation of the self-assembly units, described by the director \textbf{n}, varies with a periodicity equal to half the helical pitch, $p/2$. In many situations, the organisation of these building blocks is constrained by the curved geometry of the confining space \cite{boulig-biochol, boulig-spherulites}. The emerging structures and associated functionalities then result from an intriguing interplay between topological constraints and free energy minimization. Moreover, these structures often derive their function from having corrugated surfaces, as seen in the structural color of jeweled beetles \cite{beetle-science, beetle-mitov}. Elucidating the underlying mechanisms of cholesteric organisation in constrained environments with deformable interfaces is key not only for understanding the complex structures of these biological systems, but also for creating functional, programmable materials \cite{bionanoreview}.

A spherical droplet of synthetic CLC, whose building blocks are rod-like molecules, is the simplest system that brings us to confront these topological questions. Generally, when a CLC is confined to a bulk sphere, the helical cholesteric arrangement has an onion-like structure, provided that the molecules are tangentially anchored to the droplet surface, \textit{i.e. planar} boundary conditions \cite{odl2, depablo-choldrops}. The continuous rotation of the director field produces a pseudo-layered structure, with a $p/2$ periodicity of \textbf{n}, but no density modulation. Each cholesteric layer can be considered, as a first approximation, as a two-dimensional spherical nematic, where the director field is disrupted by the presence of unavoidable topological defects. Such topologically required defects are seen in everyday examples: for instance, drawing lines of longitude or latitude on a globe must necessarily create two defects at the poles, points where the lines are ill-defined \cite{rdkrmp}.  The spherical packing of cholesteric layers yields the formation of bulk defects, most commonly, two intertwined line defects or disclinations that span the droplet radius \cite{Sec2012}.  

Recent results have shown that imposing normal molecular anchoring at the droplet surface, \textit{i.e. homeotropic} boundary conditions, leads to more complex molecular organisations and new types of bulk defects, such as point defects or \textit{hedgehogs}, closed loops \cite{homeotdrop}, knots \cite{zumerknot} and Skyrmions \cite{dropskyrm}. Although the structures produced in CLC droplets provide opportunities for designing new functional materials, such as tunable omnidirectional lasers \cite{lasmusevic}, they still do not approach  the complexity observed in some biological systems. This is due to i) the strong anchoring conditions imposed and ii) the important role of bulk effects in droplets, which prevent the formation of complex molecular patterns on the droplet surface.

In this paper, we explore the effect of greater confinement and moderate anchoring strength using cholesteric \textit{shells}, where the CLC is confined between an inner and an outer spherical boundary. Liquid crystal shells have received special attention in the last years because of the fascinating physics emerging from their topology and their potential applications in photonics \cite{tll-nature, tll-review, lagerwall-review}. In particular, CLC shells have recently been studied under strong planar boundary conditions \cite{lagerwall-security, odl-cholshells, pH-cholshells}, revealing new intricacies in the structure of cholesteric bulk defects \cite{depablo-cholshells, alex-waltz, alex-defects}. Here, we show that moderate homeotropic anchoring results in a series of new states, where stripes form organized patterns on the shell surface. We report hierarchical structures where thin striped patterns develop within orthogonal thicker striped patterns, as well as frustrated hexagonal lattices resulting from the packing of double-spiral domains. The complexity of these meta-stable states is controllable with two parameters: the alignment strength of the CLC molecules perpendicular to the CLC-water interfaces and the confinement ratio $c = h/p$, where $h$ is the thickness of the shell. By dynamically varying these parameters we are able to induce topological transitions between states, triggering abrupt changes in the stripe widths and interconversion between \textit{nematic} defects, points where \textbf{n} is not defined, and \textit{pitch} defects, points where the cholesteric helical axis is not defined. As we will discuss in the following, the defects in cholesterics are profoundly different from defects in nematics, in particular, the transitions from one topological class to another is still not completely understood \cite{geomchol}. Because the theory of cholesteric topological defects is yet to be completely formulated \cite{geomchol}, we probe them by \textit{transitioning} between pitch and nematic defects. We simulate CLC shells with varied anchoring strength on the boundaries to analyze our findings. Better understanding and manipulation of structures in these chiral systems advances our knowledge of low-dimensional topology, structures found in biology, and self-assembly in general, broadening the functionality of CLCs in optics and as mediums for material fabrication.

\section*{Modulating anchoring at liquid crystal interfaces}

Employing microfluidics, we fabricate water-CLC-water double emulsions, where an aqueous inner droplet is coated by a CLC shell that is, in turn, dispersed in an aqueous phase (see SI, Materials and Methods). The CLC phase is composed of a mixture of 4-cyano-4'-pentylbiphenyl (5CB), nematic at room temperature, and a chiral dopant (S)-4-cyano-4-(2-methylbutyl)biphenyl (CB15). The pitch $p$ is determined by the amount of dopant present in the solution \cite{cb15pitch}. The thickness of the shell can be kinetically tuned via osmotic annealing \cite{alex-waltz, osswell}: increasing the molar salt concentration in one of the aqueous phases creates an osmotic pressure difference through the CLC shell, resulting in a flux of water across the permeable shell.  Due to the density mismatch between the inner and middle phases, the inner droplet either floats or sinks inside the outer one, leading to  a non-uniform shell thickness. This thickness heterogeneity can be reduced using heavy water (D$_{2}$O) to approximately match densities. A 1\% wt polyvinyl alcohol (PVA) in water solution is added to the inner and outer phases to stabilize both CLC-water interfaces \cite{tll-sds}. Water and PVA are also known to induce degenerate planar anchoring on the 5CB-water interface \cite{abbsurf}.

The pitch axis of a CLC orients perpendicular to a boundary with degenerate \textit{planar} anchoring so as to satisfy, without frustration, both the boundary condition and the tendency of the CLC to twist. However, at a \textit{homeotropic} boundary, there is no direction in which the pitch axis can orient such that the director is perpendicular to the entire boundary --- a homeotropic boundary competes with the tendency to twist. To study these two cases and their relationship, we induce an anchoring transition at the CLC-water interfaces using two methods: by introducing high concentrations of surfactant to the water phases around the shell or by slowly heating the shell towards the CLC-isotropic transition. Our results do not depend on our method of surface modification, leading us to conclude that the observed textures and transitions are robust.

\begin{figure*}[ht!]
\centerline{\includegraphics[width=1\textwidth]{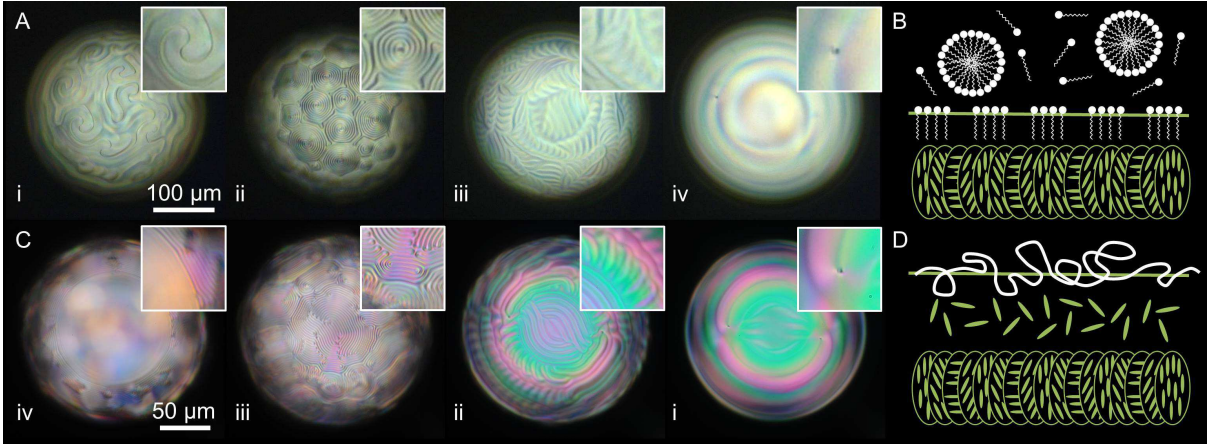}}
\caption{\label{AnchChange} \textbf{Stripe patterns emerge on a CLC shell as the anchoring is changed from homeotropic to planar.} Throughout, the pitch is 5 $\mu$m (2.8\% CB15). A) Time series of a CLC shell losing surfactant from its interface from i) to iv) (See SI Video 1). As the surfactant surface concentration is diluted from 10~mM SDS with time, the homeotropic anchoring strength weakens. Four distinct patterns, from strongest to weakest homeotropic anchoring, are identifiable: subsurface defect lines (i), double spiral domains (ii), thick planar stripes (iii), and planar anchoring (iv). B) Schematic of surfactants on the CLC-water interface. The surfactant tails cause the CLC molecules to prefer perpendicular alignment to the interface. C) Time series of a heated CLC shell, with temperature increasing from i) to iv) (See SI Video 2). A few tenths of degrees below the clearing point, an isotropic layer nucleates at the CLC-water interface, inducing weak out-of-plane anchoring of the remaining CLC. As temperature is increased, the isotropic layer grows inwards, confining the remaining CLC until the entire thickness turns isotropic. Four patterns are also identified: planar anchoring (i), thick planar stripes (ii), thin stripes (iii), and (iv) complete untwisting. D) Schematic of the isotropic layer growing between the bulk CLC and the PVA-stabilized interface between water and liquid crystal. The radius (R) and thickness (h) of the shells are R $\sim$ 150 $\mu$m and h $\sim$ 50 $\mu$m in A) and R $\sim$ 75 $\mu$m and h $\sim$ 15 $\mu$m in C).} 
\end{figure*}

In the first modality, we add sodium dodecyl sulfate (SDS) to induce an anchoring transition from planar to homeotropic (Fig.~\ref{AnchChange}A and B). This method has proven to be effective in nematic LC-water interfaces on flat surfaces \cite{abbsurf} and shells \cite{tll-sds}. We find stripe patterns that are stable over weeks (Fig.~\ref{AnchChange}A)  at  concentrations of SDS  well above the critical micelle concentration (cmc) ($>\!\!5$ mM SDS). We add sodium chloride (NaCl) to the surfactant solution in order to provide excess electrolytes that both lower the cmc and screen the electrostatic interactions between the surfactant head groups for denser packing of surfactants on the CLC-water interface \cite{abbsurf}. Osmotically swelling the emulsions additionally appears to change the amount of surfactant on the inner shell surface [evident from the pattern changes on the shell (SI Fig.~\ref{innerFCDs})], allowing for tuning of the surfactant density on the inner shell surface. Sample containers are sealed to prevent the evaporation of water so that surfactant concentrations can be kept constant over long time periods. 

Alternatively, raising the temperature of a shell with planar boundary conditions induces similar patterns (Fig.~\ref{AnchChange}C) to those seen with surfactants when the temperature reaches a few tenths of degrees Celsius below the phase transition temperature to the isotropic state. This temperature-induced anchoring transition is likely linked to the nucleation of a new interface between the PVA and the CLC phase: an isotropic-CLC interface that is known to induce weak out-of-plane molecular anchoring \cite{nem-iso}. The coexistence of a thin isotropic film with a bulk liquid crystal phase and its growth as a function of temperature has been previously reported \cite{nature-interface, yang-interface, kleman-interface, lagerwall-interface}. At the water-CLC interface, the CLC molecules interact with the amorphous, randomly-conformed PVA, creating a greater degree of disorder in the CLC near the interface than in the rest of the bulk. If the system is heated with a low rise temperature ramp (0.01$^{\circ}$C/min), isotropic domains nucleate at the interfaces before the phase transition is triggered in the bulk. These domains grow and coalesce at the interfaces, replacing the water-CLC interface (planar anchoring) by an isotropic-CLC interface (out-of-plane anchoring), as schematically represented in Fig.~\ref{AnchChange}D. As the temperature increases further, the isotropic layers expand into the shell, thinning the CLC further until the entire thickness is in the isotropic phase (Fig.~\ref{AnchChange}C and SI Videos 2 \& 3). Using two independent mechanisms to change the boundary conditions allows us to cross-corroborate our results. Indeed, the coincidence of the observed patterns establishes that these structures are truly an anchoring effect and do not result from variations in elastic constants (since the surfactant studies are at constant temperature) or molecular chemistry (since the temperature modality does not require concentration changes).  

Under planar anchoring conditions, the CLC shell has topologically required defects (Fig.~\ref{AnchChange}A-iv and Fig.~\ref{AnchChange}C-i), as stated by the Poincar\'e-Hopf theorem \cite{poincare, hopf}. Namely, the molecules cannot be aligned over the entire surface of the sphere, yielding defects whose topological charges or winding numbers --- a measure of the rotational distortion that they impose on the director field around them --- add up to $+2$. Due to the periodicity imposed by the cholesteric pitch, the CLC can be seen as a layered system, forcing the defects in the shell to extend into the bulk as defect lines \cite{depablo-cholshells, alex-waltz, alex-defects}. Alternatively, under strong homeotropic anchoring conditions, the CLC shell has disclination lines in the bulk, away from the interface (Fig.~\ref{AnchChange}A-i, SI Fig.~\ref{StrHomeoAnc}). The CLC is forced to twist rapidly at these defect lines to satisfy the energetic tendency of the system to twist while keeping the system from violating the homeotropic boundary condition \cite{homeotdrop, zumerknot}. 

Between these limiting cases, we find new states with \textit{moderate} homeotropic anchoring strengths. When a shell having initially strong homeotropic anchoring (10 mM SDS, 0.1 M NaCl, and 1\% wt PVA) is placed in a solution lacking surfactant, the surfactant surface density on the interface decreases and the shell pattern changes in a sequence shown in Fig.~\ref{AnchChange}A and SI Video 1. We find first a half-pitch-wide, thin stripe pattern (Fig.~\ref{AnchChange}A-ii). In this state, the shell surface is tiled with double spiral domains. Then, we observe a pattern with much thicker stripes (Fig.~\ref{AnchChange}A-iii), where long stripes are modulated by shorter, perpendicular stripes. Similar patterns are seen when a shell having initially planar anchoring (Fig.~\ref{AnchChange}C-i) is subjected to a linear increase in temperature, starting a few tenths of degrees below the isotropic-CLC phase transition temperature ($\sim35.3^{\circ}$C for 5CB doped with $2.8$\% CB15), at a rate of 0.01$^{\circ}$C/min (SI Video 2). As the water-CLC interface is replaced by a water-isotropic LC interface, we find a thick stripe pattern (Fig.~\ref{AnchChange}C-ii), followed by a half-pitch-wide, thin stripe pattern (Fig.~\ref{AnchChange}C-iii). As the shell is heated further, isotropic regions confine the shell to such an extent that the CLC cannot twist and is essentially nematic (Fig.~\ref{AnchChange}C-iv, inset shows nematic-CLC boundary). Further heating induces a transition to the isotropic phase.

We observe a clear dependence of the stripe patterns on the shell confinement. Being density-matched, the shell in Fig.~\ref{AnchChange}A is thick ($c\gg 1$) everywhere, while the shell in Fig.~\ref{AnchChange}C is very heterogeneous in thickness. In the latter case, the density of the CLC is greater than that of the inner aqueous droplet, causing the inner droplet to rise and thin the top region of the shell ($c \lesssim 1$). This has an effect on the observed patterns: thin stripes wrap into double-spiral regions (Fig.~\ref{AnchChange}C-iii, inset) in the thick part of the shell, as in Fig.~\ref{AnchChange}A-ii, but not in its thinnest part. Also, thicker stripes have secondary, perpendicular stripes (Fig.~\ref{AnchChange}C-ii, inset) in the thick part of the shell, as in Fig.~\ref{AnchChange}A-iii, but not in the thin part. 

How do the topology, the anchoring strength, and the shell thickness govern the formation of stripes and defect structures? To gain further insight, we use a $\mathbf{Q}$-tensor based, Landau-de Gennes model, where in the uniaxial limit, the components $Q_{ij}$ correspond to the director components $n_i$ by $Q_{ij} = 3S(n_i n_j - \delta _{ij} /3) / 2$, and $S$ is the Maier-Saupe order parameter \cite{lasso, z-rav, ms} (see SI, Materials and Methods). We vary the homeotropic and planar anchoring strength at the shell boundaries, along with the shell thickness, to study the effect of changing the boundary conditions on the system's metastable states. 

\section*{Thick stripes - the ``bent'' state}

\begin{figure*}[ht!]
\centerline{\includegraphics[width=1\textwidth]{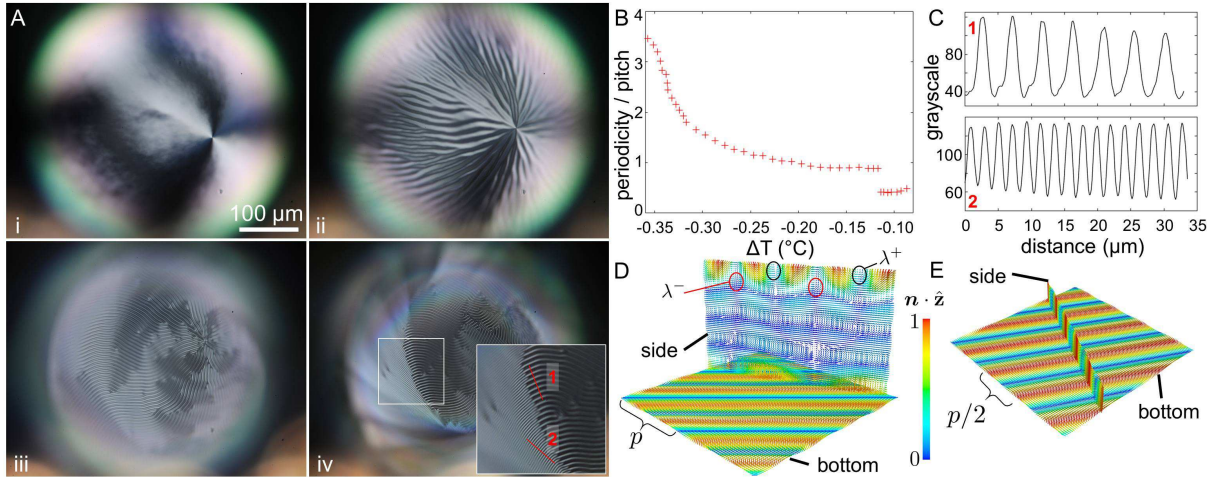}}
\caption{\label{UState} \textbf{Unbending transition under increasing confinement} A) Micrographs show a CLC shell with initially planar anchoring (i) heated towards the isotropic phase transition temperature with concomitant thinning of the cholesteric shell and introduction of weak out-of-plane anchoring. As the isotropic layer expands, thick planar stripes form initially (ii) and continuously decrease to the size of the pitch (iii), at which point, the stripe narrows discontinuously to the size of a half-pitch (iv), until the entire shell transitions to the isotropic phase (see SI Video 3). The pitch is 5 $\mu$m, the radius (R) and thickness (h) of the shell are R $\sim$ 200 $\mu$m and h $\sim$ 1 $\mu$m. B) Stripe periodicity over pitch versus temperature difference to the phase transition temperature. A continuous decrease followed by an abrupt halving is clearly seen. C) Grayscale values of stripes in (A-iv) before (red line, 1) and after (red line, 2) the front of the unbending transition show the halving of the stripe periodicity. D) Simulation of a $0.81 \times0.81\times0.54$~$\mu$m CLC slab with a 0.28 $\mu$m pitch, moderate homeotropic anchoring ($W_0 \approx 8 \times 10^{-5}~ \mathrm{J}/\mathrm{m}^2$) on the top and bottom, and periodic boundary conditions in the horizontal directions.  The outermost cholesteric layers are bent, yielding a stripe periodicity greater that the pitch. Alternating $\lambda^{+}$ and $\lambda^{-}$ pitch defect lines are labeled. E) When the slab thickness decreases from 0.54 to 0.13 $\mu$m ($\sim$half-pitch), the CLC unbends and runs parallel to the interface, creating half-pitch-wide stripes.
}
\end{figure*}

Thick stripes, stripes larger than the half-pitch, occur when the cholesteric layers are concentric in the bulk (with the pitch axis oriented in the radial direction) and distort only close to the interfaces to accommodate the boundary conditions: out-of-plane anchoring at the outer shell interface relaxes via a cholesteric texture seen in Fig.~\ref{UState}A-ii and Fig.~\ref{UState}A-iii, over an anchoring penetration depth governed by the pitch \cite{pdepth}. The layers beyond the reach of this anchoring length have planar alignment, with a pitch axis oriented perpendicular to the interface. Since the outermost layer has tilted anchoring cues from the interface, the pitch must reorient near the surface, producing a bending in the cholesteric layer closest to the boundary. Simulations of a $0.81 \times0.81\times0.54$ $\mu$m CLC slab with a 0.28 $\mu$m pitch, with moderate homeotropic anchoring on the top and bottom boundary ($W_0 \approx 8 \times 10^{-5}~ \mathrm{J}/\mathrm{m}^2$) and periodic boundary conditions in the horizontal directions reveals the director field structure of the thick stripes. This structure is depicted in Fig.~\ref{UState}D, which shows a color map of the vertical component $\mathbf{n} \cdot \hat{\mathbf{z}}$ of the director. The CLC adapts to the anchoring conditions and the rigidity of the boundary by bending the outermost cholesteric layer, as predicted by Saupe \cite{saupe}, satisfying the homeotropic anchoring near the interface and the planar anchoring from the bulk layers. These bent layers form pitch-wide stripes, consistent with experiments. In bending the layers, the system introduces pairs of defect lines with singularities in the pitch axis but smooth in the nematic director, so called $\mathbf{\lambda}^{+}$ and $\mathbf{\lambda}^{-}$ defect lines \cite{pdepth, saupe, Sec2012}, as illustrated in Fig.~\ref{UState}D. A free interface makes the stripe periodicity more pronounced (see Fig.~\ref{Coexistence}).

The stripe pattern evolves with the length ratio $c$, as shown in Fig.~\ref{UState}A, where a uniformly thin CLC shell with $c \sim 1$ undergoes a temperature-induced anchoring transition. A video of this process, showing the precise temperatures of each stage, is in SI Video 3. The shell has planar anchoring defects (Fig.~\ref{UState}A-i) initially, until it exhibits stripes a few times wider than the pitch (Fig.~\ref{UState}A-ii). As the shell is heated further, the isotropic domains at the interface grow into the shell, thinning the CLC and continuously decreasing the stripe width (Fig.~\ref{UState}B) until it is equal to the pitch (Fig.~\ref{UState}A-iii). At that point, as the system temperature is increased slightly more, the stripe width discontinuously jumps to a half-pitch (Fig.~\ref{UState}A-iv, Fig.~\ref{UState}B). A plot of the grayscale values of the stripes across the front of the periodicity change further corroborates the sudden ``doubling'' of stripes (Fig.~\ref{UState}C). Although $p$ slightly increases with temperature (in the order of 0.01 $\mu$m/$^{\circ}$C), this effect is too small to contribute to the evolution of the patterns \cite{pakula2015}.

The abrupt ``doubling" of stripes is a clear manifestation of the unbending of the cholesteric layers.  As the CLC is confined with increasing temperature, the CLC loses the bulk, concentric layers (blue layers in Fig.~\ref{UState}D for the slab geometry) that intervene between the bent layers at the interface. As the thickness decreases, the surface anchoring dominates the energetics, rearranging the director field to eliminate the $\mathbf{\lambda}^{+}$ and $\mathbf{\lambda}^{-}$ defect lines. When the slab thickness is about the size of the bent layer (about a half-pitch), the slab takes on a uniform cholesteric texture with the pitch axis parallel to the bounding surfaces, shown in Fig.~\ref{UState}E.

Varying the confinement length scale can also trigger the growth of secondary patterns. In particular, $c$ governs the presence of perpendicular sub-stripes on the shells. This effect is apparent in Fig.~\ref{AnchChange}C-ii, on the sides of the shell, where the thickness of the shell increases, and is evident in more homogeneously thick shells (Fig.~\ref{AnchChange}A-iii). The stripes may form as a result of an instability in thick shells, with a periodicity possibly set by a similar mechanism seen in hybrid nematic pancakes \cite{oksana}. The anchoring conditions between the shells and the nematic pancakes are similar: perpendicular stripes only appear when there are enough cholesteric layers to reinforce planar anchoring, providing overall hybrid anchoring for the outermost layer.  

\section*{Double-spiraled thin stripes - the cholesteric focal conic domain}

\begin{figure*}[ht!]
\centerline{\includegraphics[width=0.6\textwidth]{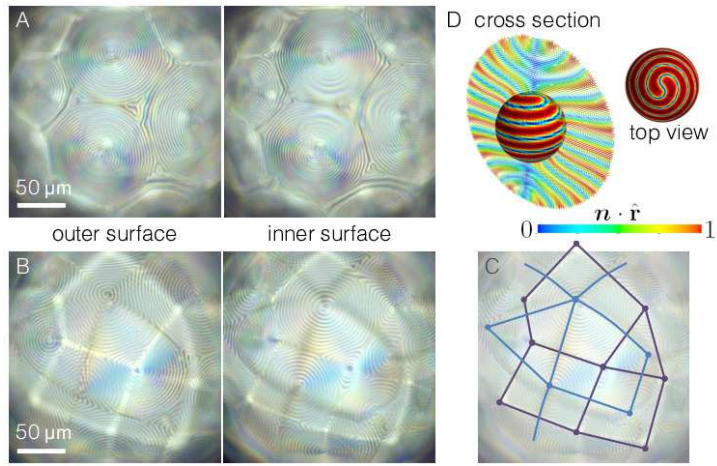}}
\caption{\label{ThinStripes} \textbf{Ordering focal conic domains on a sphere.} CLC shells with 7 mM SDS in the outer phase, without (A) and with (B) 7 mM SDS in the inner phase have thin stripes that form double spiral domains. Left micrographs focus on the outer surface of the shell, right micrographs focus on the inner surface. The pitch is 5 $\mu$m.  C) Blue lines represent the edges of FCDs on the outer surface of (B), while purple lines represent the edges of FCDs on the inner surface. The cholesteric polygonal texture comes from the staggered packing of FCDs. D) A simulated CLC shell (2.2 $\mu$m diameter, 0.70 $\mu$m thickness, and 0.42~$\mu$m pitch) with matching homeotropic anchoring conditions ($W_0 \approx 1.5 \times 10^{-5}~ \mathrm{J}/\mathrm{m}^2$) on the inner and outer surfaces. The top view shows the double spiral pattern of the outer shell surface, while the cross section additionally shows the patterning of the inner shell surface. The shell has no defect in $\mathbf{n}$, only two defects in the pitch axis, located underneath the double spirals.}

\end{figure*}

Thin stripes occur when the pitch axis is in the tangent plane of the interface, producing half-pitch-wide stripes. If the shells are thick and the homeotropic anchoring energy is sufficiently high, the thin stripes wrap into double spiral domains, as shown in Fig.~\ref{ThinStripes}. Double spirals are signatures of focal conic domains (FCDs), which are regions where the cholesteric layers bend around two focal lines \cite{pieranski, boulig-spherulites}. In this state, all the cholesteric layers across the shell thickness bend to become orthogonal to the boundaries, not merely the boundary layer (see schematics in SI Fig.~\ref{FCDDiagram} and Fig.~\ref{ThinStripes}D). The handedness of the double spirals appearing in the shells coincides with the handedness of the CLC.

The double spiral structure has been shown to arise from a locally curved surface cutting out the bent layers of the CLC \cite{boulig-spherulites, pieranski, pdepth}. In order for double spirals to be energetically favorable, the CLC shell must be thick enough such that the bulk elastic energy can overcome surface tension and \textit{deform} the shape of the interface into hills and valleys, with each hill accommodating a double spiral \cite{pieranski, pdepth, mitov}. This is evident in SI Fig.~\ref{BentStateFCD}, which shows a CLC shell with a top, thin region exhibiting the ``bent''-state discussed above (SI Fig.~\ref{BentStateFCD}, left) and a bottom, thicker region with double spiral domains (SI Fig.~\ref{BentStateFCD}, right). Moderation of the anchoring strength and lowering the surface tension of the CLC interface by the introduction of either surfactant or isotropic phase facilitate the controllable formation of these patterns that are commonly seen in nature, such as on the iridescent shell of jeweled beetles \cite{beetle-science, beetle-mitov}.  The surface corrugations at moderate anchoring strengths are related to the variability of the molecular orientation at the interface that induce gradients in the surface tension, leading to interface undulations, as seen in our simulations of the CLC-isotropic interface in Fig.~\ref{Coexistence} and in previous work \cite{cholisointerface}.

The strong confinement and spherical curvature imposed by the shell impacts the organization and shape of the FCDs, leading to interesting effects. Shells with hybrid anchoring (7 mM of SDS only in the outer phase) have double spirals that pack in hexagons and pentagons, as shown in Fig.~\ref{ThinStripes}A.  Typically, cholesteric FCDs are hexagonal in order to regularly tile planar surfaces. However, hexagons cannot pack efficiently on spheres, forcing the formation of pentagonal FCDs: the combination of hexagons and pentagons provides a full tiling of the shell surface, as seen also in fullerenes and soccer balls. Confinement also entails some anchoring violation on the inner surface. The CLC shells, with a thickness of approximately 20-30 $\mu$m and a pitch size of 5 $\mu$m, have double spiral domains with a radius of around 33 half-pitch-wide stripes, giving a total radius of around 16 cholesteric layers (about  80 $\mu$m). These layers cannot fit into the shell thickness without creating energetically costly layer distortions -- the system compromises by violating the inner shell planar anchoring, as we can see from the stripe pattern on the inner shell surface (Fig.~\ref{ThinStripes}A, right).

On the other hand, in emulsions with matching homeotropic anchoring (7 mM of SDS in the inner and outer phases), the FCDs form from both the inner and outer shell surfaces, as shown in Fig.~\ref{ThinStripes}B. The FCDs formed on the inner shell surface are mirror images of those appearing on the outer one (they show opposite handedness when observed on the microscope), but their centers are shifted. As a result, the FCDs have more varied polygonal shapes since the focal-line intersections on one surface correspond to the centers of polygons on the opposite surface, as demonstrated in Fig.~\ref{ThinStripes}C, leading to the classic staggered packing of polygonal domain textures \cite{boulig-pt2-polyg}. Blue lines represent the edges of FCDs on the outer surface (Fig.~\ref{ThinStripes}B, left), while purple lines represent the edges of FCDs on the inner surface (Fig.~\ref{ThinStripes}B, right). 

We observe that greater homeotropic anchoring strength is needed to induce FCDs from the inside of CLC shells than from the outside for shells with $c\gg 1$ because the creation of hills is less costly at the outer surface. This is also an effect of curvature: because the FCDs formed on the outer and inner shell surfaces are mirror images, the hills formed on the outer surface have the same type of curvature as that of the surface that they deform, while those formed on the inner surface have the opposite curvature, and then, are less energetically favorable. Adding SDS to the inner water phase alone, no matter the concentration, fails to form double spiral domains on the inner shell surface. To induce FCDs to form \textit{only} on the inner shell surface, an extreme salt concentration (1 M NaCl) must also be added to the inner water phase, as shown in SI Fig.~\ref{innerFCDs}. The high salt concentration allows the surfactants to pack more densely on the surface, effectively increasing the homeotropic anchoring energy of the system. 

Similar patterns are observed in a thick CLC shell with the temperature anchoring transition. When the thin-stripe pattern emerges, hexagonal FCD packing is observed before the polygonal texture appears (SI Video 2). At the beginning of the temperature ramp, the isotropic layer nucleating at the shell interfaces is very thin, and the FCDs form just on the outer shell surface where it is easier to deform the interface. As the temperature increases, the isotropic region increases in thickness, so that the CLC  bulk elastic energy can overcome the now weaker isotropic-CLC interfacial tension on both sides of the shell, allowing for a polygonal texture to form. 

We probe the director structure of FCDs via simulation. In Fig.~\ref{ThinStripes}D, the director structure of a system with a 2.2~$\mu$m diameter, a 0.7~$\mu$m thickness, and a $0.42$~$\mu$m pitch ($c\approx 1.7$) is plotted with a color map of the radial component $\mathbf{n} \cdot \hat{\mathbf{r}}$ of the director. Note that the length scales probed by the simulations are limited by the mesh spacing of the simulation grid, although we do not expect significant differences in larger shell simulations, especially in regions away from the core of the pitch defects (see SI, Materials and Methods for more information). For moderate homeotropic anchoring conditions ($W_0 \approx 1.5 \times 10^{-5}~ \mathrm{J}/\mathrm{m}^2$), the CLC twists along the surface of the sphere, producing two FCDs at the poles of the sphere. The spherical topology of the system dictates a minimum of two focal conic domains, a state found in experiments when the CLC shells are left to anneal for about one month.
 The surface stripes of the FCDs end at double spirals (Fig.~\ref{ThinStripes}D, top view), as they do in experiments. The cross section reveals that the cholesteric layers underneath the double spirals bend upwards. The system does \textit{not} have any director defects - only \textit{pitch} defects, areas with an ill-defined twist direction. The pitch defect of the FCD is evident from the cross section in Fig.~\ref{ThinStripes}D, where alternating red and blue values of $\mathbf{n} \cdot \hat{\mathbf{r}}$ along the spherical surface (indicating a twist axis along the sphere surface) collide with a region at the poles where the twist axis is pointing radially (the up-down direction in the figure), indicating a discontinuity in the pitch axis. The pitch defect appears to consist of two intertwined $\lambda$-lines, similarly to the defect core discussed in Ref.~\cite{pieranski}.  Note that unlike a typical focal conic domain at a single free interface where the twisted $\lambda$ lines connect and terminate in the bulk \cite{pieranski}, the lines in the shell in Fig.~\ref{ThinStripes}D appear to span the entire shell thickness, generating a double spiral at the inner shell surface, as well.   

\section*{Transition back to planar from homeotropic equilibrium}
A CLC shell with multiple double spiral domains is in a metastable state: after about one month, the FCDs coalesce with one another and only the two, topologically required FCDs remain at opposite poles on the sphere, similar to simulation results in Fig.~\ref{ThinStripes}D. However, unlike simulation results, the majority of shells with only two FCDs have double spirals of \textit{opposite} handedness. This is apparent in Fig.~\ref{ThinStripes}A when comparing the handedness of double spirals of the right and left panels, where the right panel is focused \textit{through} the shell. Over long time periods, FCDs are formed on both sides of the shell because of the solubility of hydrocarbon surfactants in 5CB. Since FCDs of a given cholesteric all have double spirals of the same handedness, one double spiral must have formed from the inner surface in order to have the opposite handedness on the outer surface. The ubiquity of shells with oppositely winding FCDs suggests that it is energetically more favorable for FCDs formed on the same side of the shell to coalesce than it is for FCDs on opposite sides.

\begin{figure*}[ht!]
\centerline{\includegraphics[width=1\textwidth]{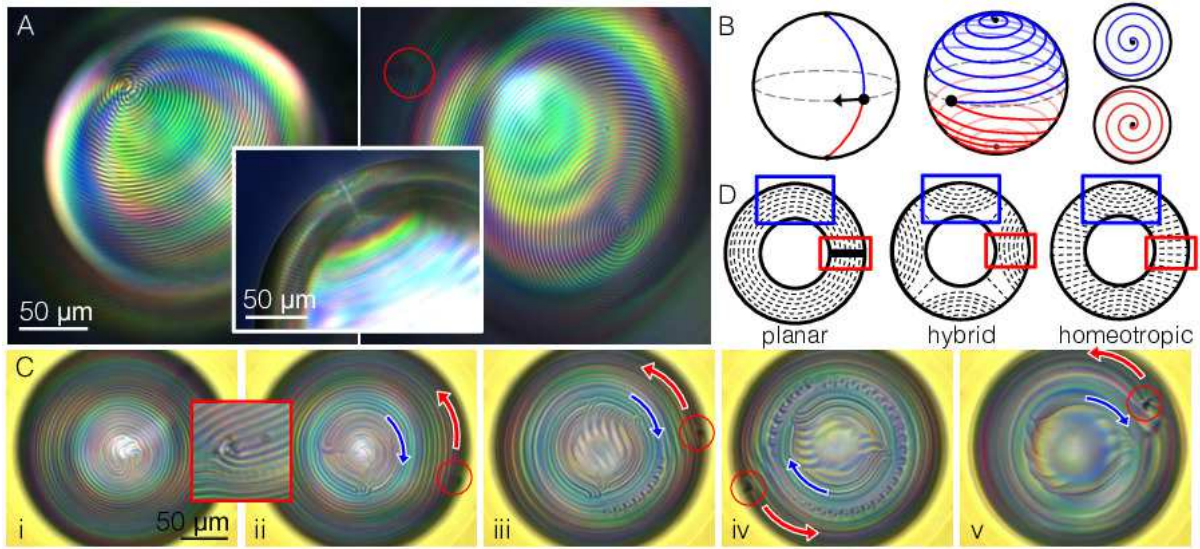}}
\caption{\label{EqThinStripes} \textbf{Equilibrium thin stripes.} A) A CLC shell with 7 mM SDS in the water phases equilibrates after one month, leaving only two, topologically required FCDs. Inset shows a side view of a FCD. B) Schematic showing how two spirals can form on a sphere from a line connecting the two poles. For two spirals of opposite handedness, a point (black) must exist where the handedness is not defined, playing the role of the stripe dislocation in the right of A) and in the red inset of C). C) The two FCDs in an equilibrated CLC shell have double spirals of the opposite handedness, requiring a stripe dislocation (red inset). When the surfactant is washed away, the stripes rotate as they become wider and unwind, evident from the opposite rotation of FCD poles (blue arrows) with respect to the stripe dislocation (red arrows). The stripes widen first at the poles (i-ii) until an instability occurs (iii-iv). The stripe dislocation becomes the planar defect. D) Schematic of cholesteric layers in a shell for planar, hybrid, and homeotropic anchoring. Blue boxes highlight that layers do not rearrange much to accommodate the anchoring change near a FCD. The opposite is true away from a FCD, highlighted by red boxes.}
\end{figure*}

The schematic in Fig.~\ref{EqThinStripes}B depicts the formation of \textit{single} spirals, for the sake of simplicity, on a sphere by taking a flexible line connecting two poles and moving the mid-point of the line around the equator until spirals are formed at the poles. The spirals that result are of opposite handedness  (Fig.~\ref{EqThinStripes}B, right) and there must be a defect where the stripe wrapping direction switches handedness. In experiments, this defect is a stripe dislocation near the shell equator, an area in which extra stripes are nucleated, disturbing the periodic stripe ordering. The stripe dislocation is circled in red in the right panel of Fig.~\ref{EqThinStripes}A. In another shell, the stripe dislocation is visible from a side-view (Fig.~\ref{EqThinStripes}C, inset).

We probe the topological defects in CLCs by \textit{transitioning} this shell's boundary conditions from homeotropic to planar to change the topology of the CLC bulk. Cholesteric topological defects are profoundly different than those in the nematic phase.  Though an analogy is commonly drawn between cholesteric ``layers'' and smectics, this is known to be a crude approximation, if at all, to the actual topology of the cholesteric phase \cite{geomchol, pieranski}. 

To illustrate this difference between defects in nematic shells and cholesteric shells, let us consider a nematic shell with planar anchoring on both boundaries.  The Poincar\'e-Hopf theorem requires that on each boundary the nematic has a total defect charge of $+2$.  If the shell is thin enough that we can approximate it as a single surface, then energetic considerations would suggest that the charge breaks up into four $+1/2$ disclinations \cite{tll-nature}.  However, as the shell thickens, point disclinations on the inner and outer surface will be connected by a bulk defect line (recall that in the nematic, bulk defect lines are characterized by a $\mathbb{Z}_2 = \pi_1(\mathbb{R}P^2)$ charge only).  At some point, the energy will be lowered by combining the bulk disclination lines in pairs, allowing them to ``escape into the third dimension,'' leaving behind charge $+1$ disclinations on the surfaces \cite{vv-nemshell}. We can alternatively consider homeotropic surface anchoring: in the case of a homeotropic nematic shell, the final low energy state has no defects {\sl whatsoever}.  If we were considering a spherical droplet, there would be a radial defect or hedgehog in the center but, because we are considering a shell, the defect is virtual.  

Now let us consider our experiments where we \textit{transition} from the original planar state of a nematic shell to the final homeotropic state. In this case, defects in the bulk coalesce and \textit{escape} as the topology of the system changes \cite{tll-sds}. The cholesteric, however, {\sl does not allow} these waltzes. In addition to the director, a cholesteric must have a pitch axis perpendicular to the director, together forming a triad at each point in space. Triads do not escape. For instance, if the director deforms to remove the winding of a $+1$ disclination line, the pitch axis will now wind --- the nematic line defect becomes a pitch line defect. We observe the reverse, a pitch defect becoming a nematic defect, when we transition the experimentally equilibrated CLC shell from moderate homeotropic to planar anchoring.

In Fig.~\ref{EqThinStripes}C, this anchoring transition reveals stripe dynamics that are associated with the pitch axis reorienting on a sphere --- from parallel to the shell surface to radial. When the surfactant is washed away from equilibrated CLC shells, the stripe pattern rotates as a result of the stripes becoming thicker and unwinding. In Fig.~\ref{EqThinStripes}C, FCD poles rotate with the opposite handedness (blue arrows) compared to the stripe dislocation (red arrows), matching the unwinding dynamics of the simplified schematic in Fig.~\ref{EqThinStripes}B --- as the stripes become wider, the single spiral will rotate with a handedness opposite that of the equatorial defect. The stripes widen sharply at the poles first, indicative of the pitch axis tilting away from parallel, causing the outer cholesteric layer to bend as they do in the ``bent''-state (i-ii). Evidently, the pitch axis reorientation is energetically more favorable near the pitch defects of the FCDs. While the stripes unwind, more layers bend and widen at the poles until the cholesteric layers are planar and concentric, apparent from the lack of stripes. During this process, an instability occurs on the thin, top portion of the shell (iii-iv) and stripes at a set distance away from the pole develop perpendicular sub-stripes. 

Near the poles where the stripes initially widen the most, the cholesteric layers are strained and start to undulate perpendicular to the stripe direction. This way of relieving the twist energy is reminiscent of the Helfrich-Hurault instability of a cholesteric frustrated by a magnetic field \cite{helfhurault}. The undulated areas widen with the rest of the stripes as the anchoring transition continues, and a defect comes into view (v). As the anchoring evolves from homeotropic to planar, new defects must form: concentric layers of cholesteric topologically require a total defect charge of +2.  We observe that these defects form precisely where the dislocation was located. The planar defect moves closer to the thinner region of the shell to reduce the elastic distortion. Videos of this process are in SI Videos 4-6. 

The pitch defects of the shell adapt to the anchoring transition by becoming sites where the boundary conditions change the most. The pitch defects of FCDs accommodate stripe width changes first, while other pitch defects, such as the stripe dislocation, become the topologically required planar defects. Fig.~\ref{EqThinStripes}D illustrates that if a defect were to form on the shell with one or both surfaces having homeotropic anchoring, the defect is more energetically likely to form near the equator, away from the double spiral poles. The textures in the blue boxes near a double spiral demonstrate that the layers need not drastically change their orientation to accommodate the anchoring change. More rearrangements are needed away from a double spiral, as highlighted by red boxes. Indeed, in the hybrid schematic in Fig.~\ref{EqThinStripes}D, the red box encompasses a pitch defect, while in the homeotropic schematic, the red box features extreme changes in director orientation compared to the planar schematic. The great degree of rearrangement needed to adapt to the anchoring transition at the equator makes a defect at the equator more probable. Energetic defects not required by the topology can become topologically required by dynamically changing the system's boundary conditions. This process further demonstrates that the direction of the pitch axis depends upon the nematic director itself. The topology of the cholesteric is not just a recapitulation of the biaxial nematic. Indeed, the theory of cholesteric defects is still not completely formulated. This transition of the equilibrated CLC shell from moderate homeotropic to planar anchoring further sheds light on the geometry and topology of CLCs.

We have demonstrated the ability to reproducibly control defect textures and patterns in CLC shells by tuning confinement and the out-of-plane anchoring strength via two methods: by varying the chemistry of the water phases and the shell interfaces and by slowly increasing the temperature of the system. We control the anchoring strength to be \textit{moderate}, allowing the CLC to twist \textit{at} the surface, better mimicking structures seen in biological systems. This corrugated surface additionally lends itself to complex assemblies at the CLC-water interface. Numerical work further corroborates our description of the nuanced complexion of stripes, defects, and topology. This comprehensive study of cholesteric textures through anchoring transitions and geometric confinement and the topologically-constrained pathways from one defect configuration to another lays the groundwork for designing LC defects as templates for nanomaterials and deepens our understanding of the cholesteric phase.

We thank D.A. Beller, M. Benzaquen, C. Blanc, S. \v{C}opar, O. Dauchot, E. Lacaze, F. Livolant, and S. \v{Z}umer, for fruitful discussions and support. This work was supported by NSF MRSEC Grant DMR1120901, by ANR Grant 13-JS08-0006-01, and by Institut Pierre-Gilles de Gennes, Program ANR-10-IDEX 0001-02 PSL and ANR-10-EQPX-31. M.O.L. acknowledges support from the US Department of Energy, Office of Basic Energy Sciences, Division of Materials Sciences and Engineering under Grant No. DE-FG02-05ER46199, and from the Simons Foundation for the collaboration ``Cracking the Glass Problem'' (Grant No. 454945). R.D.K. was partially supported by a Simons Investigator grant from the Simons Foundation.

\newpage

\renewcommand\thefigure{S\arabic{figure}}    
\setcounter{figure}{0} 

\section*{Supplementary Information}

\section*{Materials and Methods}

\paragraph*{Reagents, Microfluidics and Optical Characterization}       For the cholesteric liquid crystal (CLC), we use 5CB (4-cyano-4'-pentylbiphenyl, Kingston Chemicals Limited and Sython Chemicals) doped with CB15 ((S)-4-cyano-4-(2-methylbutyl)biphenyl, EMD Performance Materials and Sython Chemicals) for a right-handed cholesteric pitch. The pitch of the CLC is determined with a Grandjean-Cano wedge cell \cite{GJwedge, Cwedge}. 2.8\% CB15 gives a pitch of $\sim$ 5$\mu$m. CLC shells are produced using glass capillary microfluidic devices, similar to those of previous works \cite{ufluidics}. 

\begin{figure*}
\centering
  \includegraphics[width=1\textwidth]{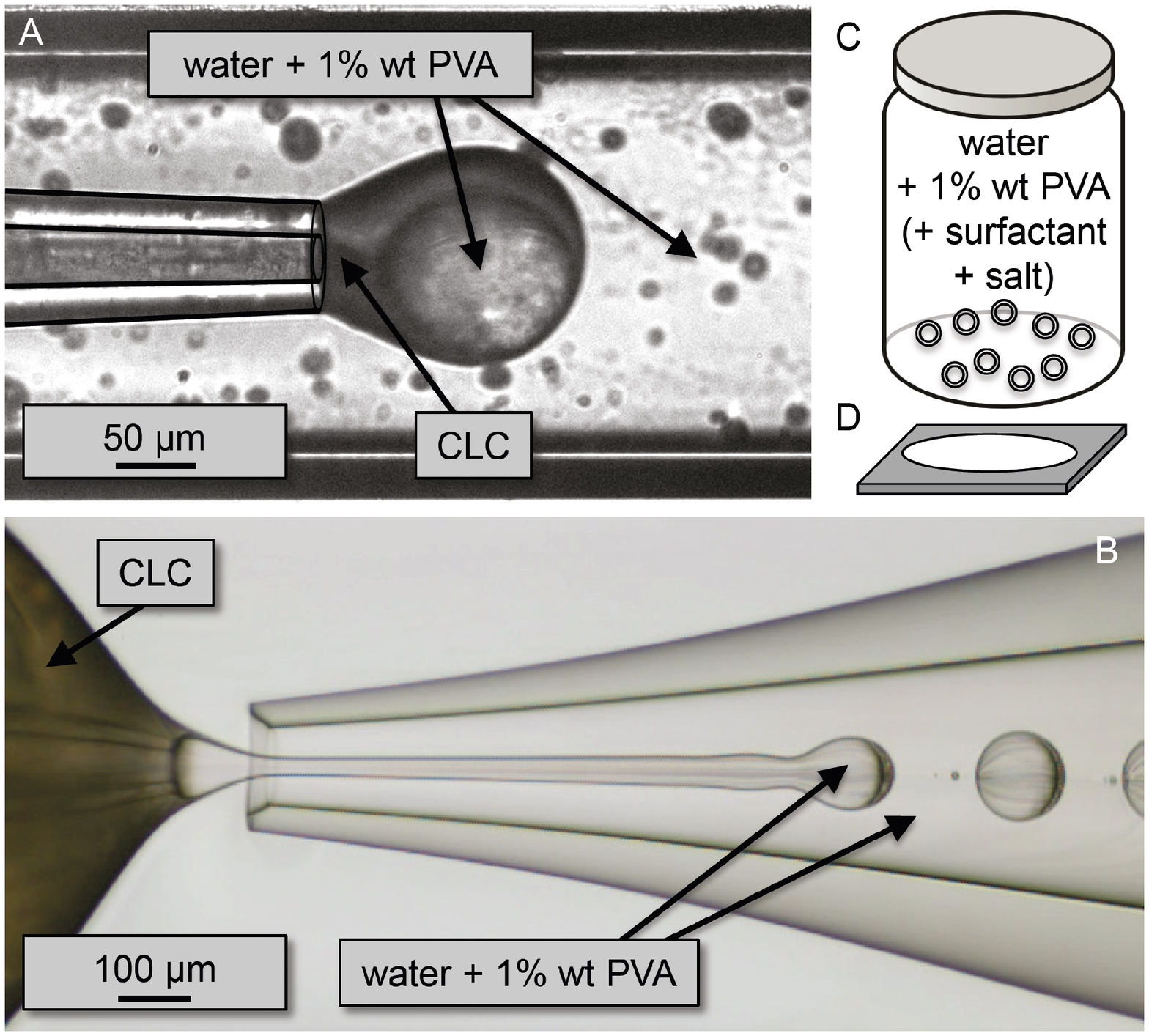}
  \caption{ \textbf{Fabrication of CLC shells.} Cholesteric liquid crystal (CLC) double emulsions are produced using either a microfluidic device with two nested coaxial capillaries, in the case of surfactant experiments (A), or a microfluidic device with two tapered capillaries facing opposite directions, for temperature experiments (B). The inner and outer phases are water with 1\% wt PVA. The middle phase is CLC. The shells are collected into a vial containing water and 1\% wt PVA (C). For surfactant experiments, the shells are then pipetted into a vial containing water, 1\% wt PVA, 0.1 M NaCl, and varying concentrations of SDS. After about 10 minutes, the CLC shells are pipetted into a viewing chamber that is then sealed to prevent evaporation (D).}
  \label{ExpSetup}
\end{figure*}

Two different geometries for devices were used, both yielding similar shells. For surfactant experiments, we utilized three nested, coaxial capillaries, as shown in SI Fig.~\ref{ExpSetup}A. The water phases have 1\% wt PVA (polyvinyl alcohol, Sigma-Aldrich, 87-89\% hydrolyzed, average $M_{w} = 13-23$ kg/mol) to stabilize the emulsions. For temperature experiments, the geometry of the device consists in two tapered cylindrical glass capillaries facing opposite directions, fitted in a square capillary, as shown in SI Fig.~\ref{ExpSetup}B. The inner water phase has 1\% wt PVA (Sigma-Aldrich, 87-89\% hydrolyzed, average $M_{w} = 13-23$ kg/mol), while the outer water phase has 1\% wt PVA and 10\% wt glycerol (VWR Chemicals), in order to increase its viscosity. Glycerol cannot be used for surfactant transition experiments, as glycerol affects the surface activity of the hydrocarbon surfactant. 

For the surfactant anchoring transition, after the double emulsions are collected, they are left to stand for a few hours to allow them to equilibrate and settle to the bottom of the vial. To induce homeotropic anchoring on the outer surface of the CLC shell, the emulsions are then pipetted into vials containing aqueous solutions of 1\% wt PVA, at least 0.1 M NaCl (sodium chloride, Fisher Scientific), and varying concentrations of SDS (sodium dodecyl sulfate, Sigma-Aldrich), in the range 0-10 mM (SI Fig.~\ref{ExpSetup}C). The vial is gently shaken and left for 10 minutes before pipetting and sealing the drops into a viewing chamber (SI Fig.~\ref{ExpSetup}D). An upright microscope in transmission mode fitted with crossed polarizers (Zeiss AxioImager M1m) and a high-resolution color camera (Zeiss AxioCam HRc) is used to take polarized micrographs. Samples are viewed over many days and weeks because the CLC relaxation time is long --- the CLC shells need many weeks to reach their equilibrium state. For emulsions observed over long time periods, heavy water (D$_{2}$O, Sigma-Aldrich) is used for approximate density matching with the CLC to prevent one side of the shell from becoming too thin. 

The role of NaCl is to increase the interfacial density of the surfactant \cite{abbsurf} and to decrease the critical micelle concentration (cmc). The cmc of SDS at 25 $^{\circ}$C in a 0.1 M NaCl solution is $\sim$ 1.47 mM \cite{sds-cmc}. The cmc of SDS decreases with increasing salt concentrations. No effect of the hydrolysis of SDS is seen on the shell pattern when comparing newly prepared SDS solutions and older solutions. A higher molar concentration of NaCl in the inner phase can additionally be used to cause the inner drop to osmotically swell with time \cite{osswell}, facilitating the study of CLC shells of varying thicknesses. The amount of water added to the inner phase after osmotically swelling the shell (determined from the volume and shell thickness) is not enough to dilute the inner salt concentration to match that of the outer water phase. Thus, we hypothesize that salt must move through the shell during the swelling process. Typically, 1 M NaCl is added to the inner water phase in experiments for osmotic swelling. 

For the temperature anchoring transition, temperature control is achieved using a mK1000 temperature controller and a TS62 thermal stage (Instec). In all experiments, the shells undergo a temperature ramp of 0.01$^{\circ}$C/min spanning several hours. Observation is conducted under an upright polarizing microscope (Nikon Ni-U) equipped with a DSLR camera (Nikon D300s).  

\paragraph*{Numerical Methods}    To simulate the cholesteric droplets, a Landau-de Gennes free energy \cite{z-rav} was minimized to find the symmetric, traceless rank-2 tensor field $\mathbf{Q}(\mathbf{x})$ describing the orientation of the liquid crystal on a cubic lattice of sites $\mathbf{x}$.  In the uniaxial limit, the director direction  $\mathbf{n}$ (expressed as a vector) is related to $\mathbf{Q}$ via $\mathbf{Q}=(3S/2)\left[\mathbf{n} \otimes \mathbf{n}-\mathbf{I}/3\right]$, $\otimes$ is a dyadic product (i.e., $[\mathbf{n} \otimes \mathbf{n}]_{ij} \equiv n_i n_j$) and $\mathbf{I}$ is the identity tensor.  The bulk free energy is given by
\begin{equation}
f_{\mathrm{bulk}}= \frac{A}{2}  \Tr \mathbf{Q}^2  + \frac{B}{3} \Tr \mathbf{Q}^3+ \frac{C}{4}( \Tr \mathbf{Q}^2)^2,
\end{equation}
where $\Tr$ is the trace.  It can be shown that a uniaxial $\mathbf{Q}$-tensor minimizes this free energy with a value of $S=S_0\equiv(-B+\sqrt{B^2-24AC})/6C$.  The  gradient contribution to the free energy reads
\begin{equation}
f_{\mathrm{grad}}= \frac{L_1}{2} (\nabla \times \mathbf{Q}+2q_0 \mathbf{Q})^2 + \frac{L_2}{2} (\nabla \cdot \mathbf{Q})^2+f_{24},
\end{equation}
where $q_0$ is $2 \pi/p$ for a cholesteric pitch $p$. We also have a saddle-splay free energy density $f_{24}$ which is written in terms of the $\bm{Q}$ components $Q_{ij}$ as $f_{24} \equiv L_{24}(\partial_i Q_{jk}\partial_k Q_{ij}-\partial_i Q_{ij} \partial_k Q_{jk})/2$, where we sum over all indices $i$,$j$, and $k$. The Landau-de Gennes free energy is minimized using the conjugate gradient method from the ALGLIB package (\url{http://www.alglib.net/}).  We can divide by an overall energy scale to set $A=-1$.  Then, to check the robustness of the simulated patterns, two sets of constants were used.   In the one-constant approximation, used for the CLC slab simulations in the main text, we chose standard values $B=-12.33$, $C=10.06$, $L_1= L_2\equiv L=2.32$, (and no saddle-splay term: $L_{24}=0$) corresponding to values in Ref.~\cite{z-rav}.  To check that the one-constant approximation, lack of a saddle-splay term, and particular values of the elastic parameters did not significantly influence our results, we also performed simulations in the two-constant approximation with $L_1 \neq L_2$.  In the spherical shell simulations in the main text we used   $B=-1.091$, $C=0.6016$, $L_1=0.00761$,  $L_2=0.02282$, and $L_{24}=L_2/2$. 

The anchoring was modeled using a Rapini-Papoular surface potential for the perpendicular anchoring and a degenerate planar anchoring potential
\begin{equation}
f_{\mathrm{hom.}} = W_0 \int \mathrm{d} A \, \Tr [(\mathbf{Q}-\mathbf{Q}^{\parallel})^2],
\end{equation}
where $\mathbf{Q}^{\parallel}= (3 S_0/2)( \bm{\nu} \otimes \bm{\nu}-\mathbf{I}/3)$ is the uniaxial $\mathbf{Q}$-tensor constructed to orient parallel to the surface normal vector $\bm{\nu}$.  That is, we penalize deviations of the $\mathbf{Q}$ tensor away from a uniaxial one oriented along the boundary surface normals. The planar anchoring condition is similar, except we penalize whenever the $\mathbf{Q}$ tensor deviates away from the plane of the surface
\begin{equation}
f_{\mathrm{plan.}} = W_1 \int \mathrm{d}A \left[ \Tr[(\bar{\mathbf{Q}}-\bar{\mathbf{Q}}^{\perp})^2] +(\Tr \mathbf{Q}^2-3 S_0^2/2)^2 \right],
\end{equation}
where $\bar{\mathbf{Q}} \equiv \mathbf{Q}+S_0 \mathbf{I}/2$ and $\bar{\mathbf{Q}}^{\perp}=(\mathbf{I}- \bm{\nu} \otimes \bm{\nu}) \bar{\bm{Q}} (\mathbf{I}- \bm{\nu} \otimes \bm{\nu})$ is a  projection of the $\bm{Q}$ tensor on the plane perpendicular to the surface normal $\bm{\nu}$.

The length scales we can simulate are limited by the mesh spacing $\Delta x$ of the cubic grid used to compute the derivatives in the elastic terms $f_{\mathrm{grad}}$.  This mesh spacing is typically set by the nematic correlation length, which describes the characteristic spatial variation of the Maier-Saupe order parameter $S$ and, therefore, the size of nematic defects \cite{z-rav}.  In the one-constant approximation we use, this spacing is given by $\Delta x =\sqrt{2K/[9 S_0^2L \tilde{A}]}\approx4.4~\mathrm{nm}$,
where $K= 10^{-11}~\mathrm{N}$ is the Frank elastic constant for 5CB, $S_0 \approx 0.533$, and $\tilde{A}=0.172 \times 10^6~\mathrm{J}/\mathrm{m}^3$ is the unscaled constant $A$ in the potential.  Choosing this lengthscale allows us to capture any nematic defects, which is especially important for strong anchoring conditions as shown in Fig.~\ref{StrHomeoAnc}.  However, in shells with weak homeotropic anchoring, we do not find any nematic defects, and we expect that the potential $f_{\mathrm{bulk}}$ is minimized at all points in the simulation.  In this case, it should be possible  to use a larger spacing $\Delta x$, and we do not expect our results to change much if we simulated larger shells.   For example, in the two-constant approximation, we tried a larger mesh spacing of $\Delta x \approx 10$~nm.  This choice did not influence the major features of our results, such as the focal conic domains in the spherical shells and the orientation of the pitch axis along the sphere surface.  We do not expect to find differences for even larger mesh spacings; the only issue is that it becomes more difficult for the simulation to find a minimum energy state because the energy becomes dominated by contributions from $f_{\mathrm{bulk}}$.

\section*{SI Section 1: Bulk defect lines with strong homeotropic anchoring}
 
\begin{figure*}
\centering
  \includegraphics[width=1\textwidth]{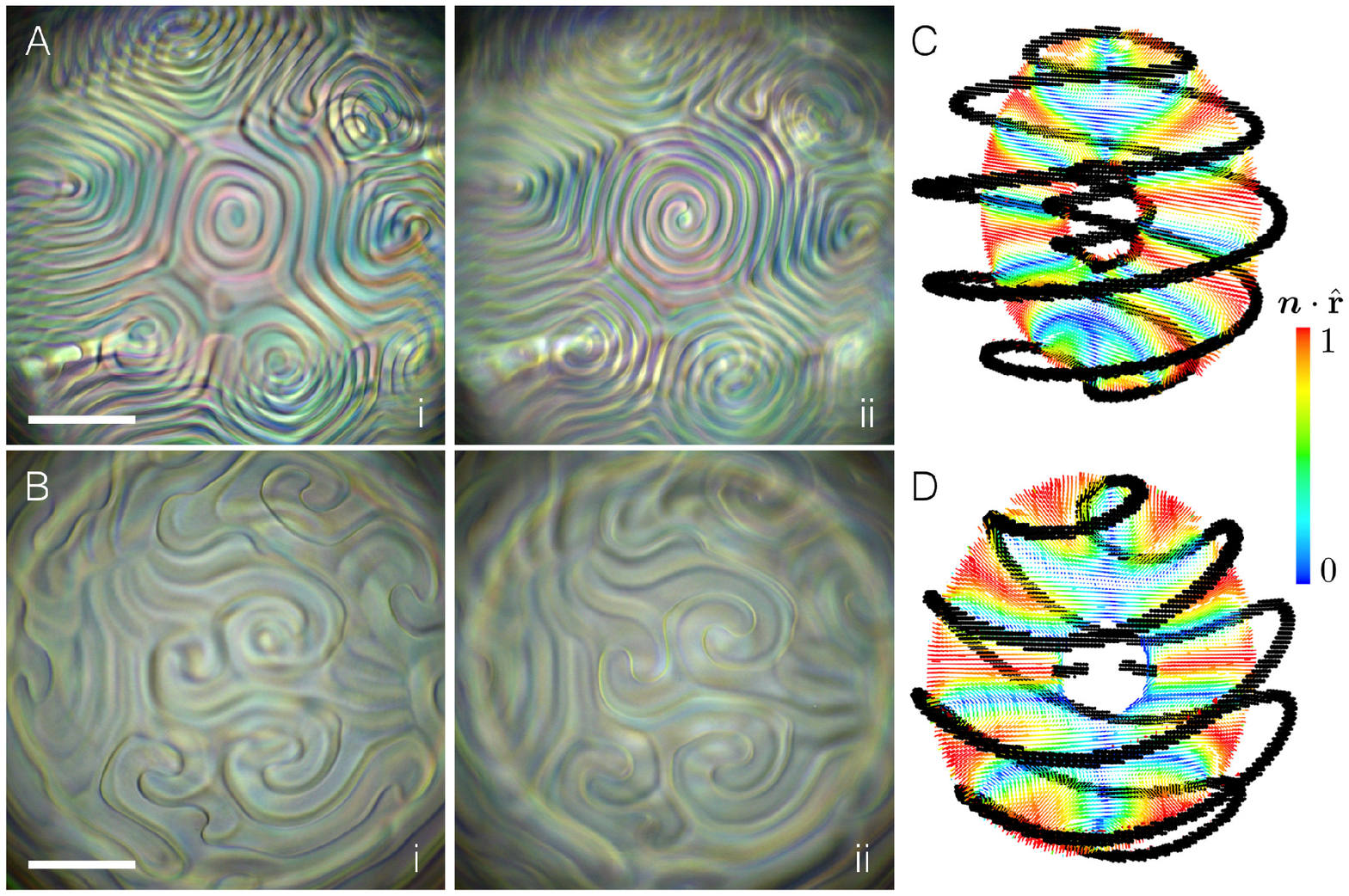}
  \caption{\textbf{Subsurface defect lines with strong homeotropic anchoring.} CLC shells with 0.1 M NaCl in the water phases and 10 mM SDS in both the inner and outer phases (A) and only in the outer phase (B) have subsurface defect lines. Pictures are focused either on the inner surface (i) or on the outer surface (ii). Scale bars are all 50 $\mu$m. A and C) Because the defect lines (black lines in the simulations) form from both surfaces, the defect lines align to reduce the distortion. B and D) Because the pitch axis must distort away from being parallel to the surface with hybrid anchoring, double spirals delineated by the defect line are less tightly wound. In the simulations in C and D), the one-constant elastic constant approximation is used. The shells have a radius of 0.42 $\mu$m and a thickness of 0.31 $\mu$m. In this case, the smallness of the simulated shell is necessitated by the presence of the defect lines at the surface, which have a size governed by the nematic correlation length, which is on the order of a few nanometers \cite{z-rav}.  The anchoring is strong, with $W_0 \approx 4 \times 10^{-3}$~J/m${}^2$ for the homeotropic surfaces and $W_1=W_2 \approx  4 \times 10^{-3}$~J/m${}^2$ for the planar ones (see SI). The pitch is 0.3 $\mu$m. }
  \label{StrHomeoAnc}
\end{figure*}
 
When the homeotropic anchoring strength is increased beyond that of the thin stripe pattern, the ground-state twist of the cholesteric is further frustrated. To fit in more LC molecules at the interface, the twist of the cholesteric is pushed away from the surface and into the LC bulk. Subsurface defect lines are formed instead of the alternating, thin homeotropic and planar stripes that have the width of a half-pitch. Strong homeotropic anchoring forces the cholesteric to be perpendicular to the entire boundary and, in order to satisfy the energetic twist of the system, the cholesteric must twist rapidly in the bulk in discrete areas, creating defects. 

Subsurface defect lines are shown in shells with matching homeotropic anchoring and hybrid anchoring in SI Fig.~\ref{StrHomeoAnc}A and C and Fig.~\ref{StrHomeoAnc}B and D, respectively. In SI Fig.~\ref{StrHomeoAnc}A and B, the left column is focused on the inner surface, while the right is focused on the outer surface. The SDS concentration is 10 mM. Similar defects are seen in previous work on single emulsions of CLCs \cite{homeotdrop, zumerknot}. 

Although defect lines are pushed away from the surface, they still remain relatively near the surface, as shown in simulations of shells with matching strong homeotropic anchoring conditions (SI Fig.~\ref{StrHomeoAnc}C, black). The defect lines formed near the inner and outer surfaces reduce the distortions from their defects further by lining up with one another, evident in both simulations and experiments (SI Fig.~\ref{StrHomeoAnc}A). Subsurface defect lines in hybrid shells, with strong homeotropic anchoring on the outer surface, have stripes with a wider width as a consequence of the cholesteric layers being further frustrated (SI Fig.~\ref{StrHomeoAnc}B and D). Simulations reveal that the inner, planar surface has defects typically associated with planar, concentric layers (SI Fig.~\ref{StrHomeoAnc}D), distorting the cholesteric layers towards the outer surface, resulting in spirals that are not as tightly wound as those in shells with matching homeotropic anchoring conditions.

\section*{SI Figures and Videos}

\begin{figure*}
\centering
  \includegraphics[width=0.9\textwidth]{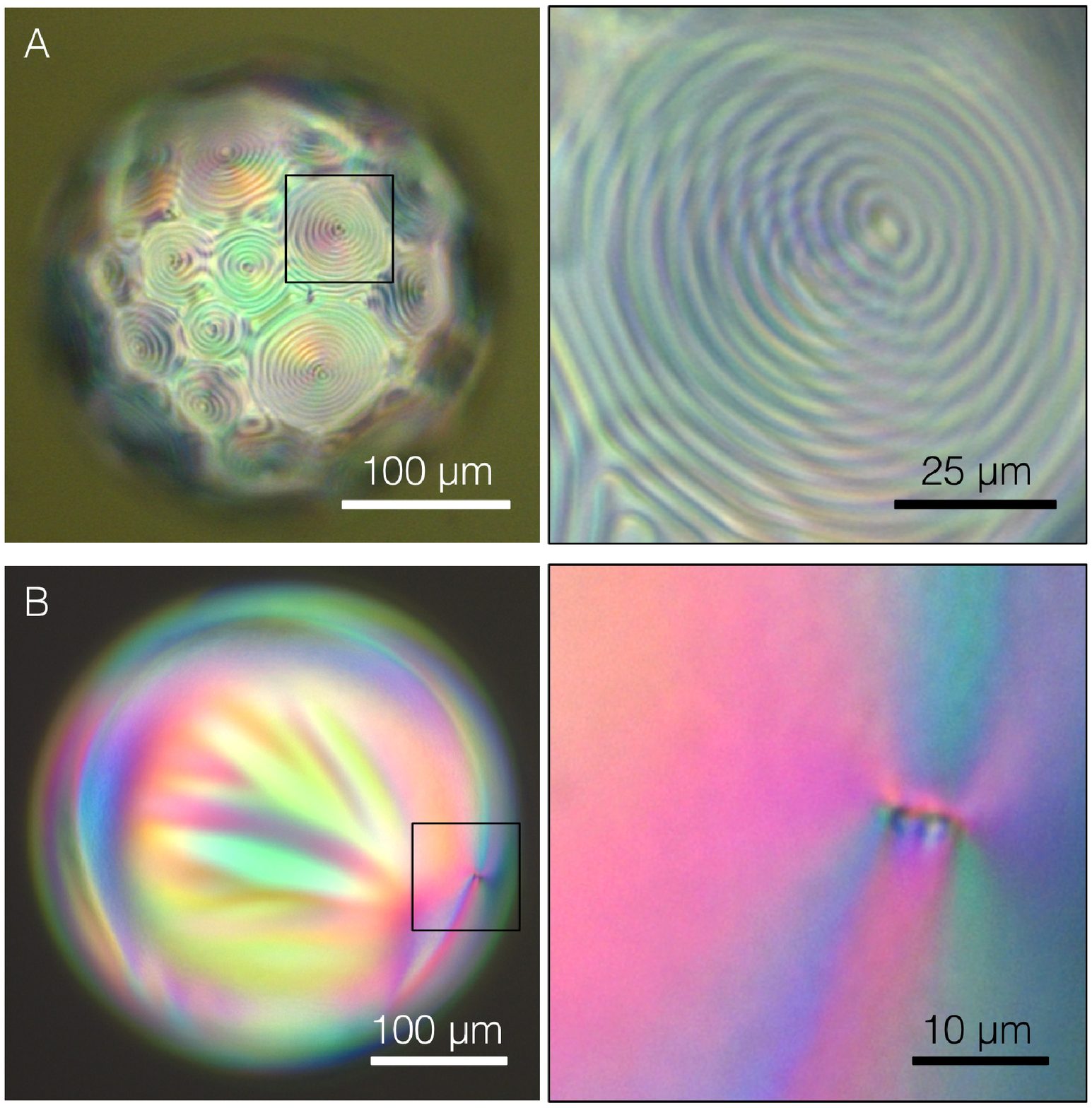}
  \caption{\label{innerFCDs} \textbf{Focal conic domains (FCDs) form on inner shell surface.} A CLC shell has 5 mM SDS, 1 M NaCl, and 1\% PVA in the inner phase water phase and 5 mM SDS, 0.1 M NaCl, and 1\% PVA in the outer phase. FCDs can only be created on the inner shell surface alone when an extreme concentration of salt (1 M NaCl) is added to the inner water phase. A) Before the shell completes osmotically swelling, FCDs, formed from the inner shell, are seen (zoomed on right). The high salt concentration screens the SDS surfactant head group, allowing the surfactant to pack more densely on the interface. B) After one day, the shell completes its swelling. The surfactant and salt concentrations are decreased, and the FCDs disappear as the inner salt concentration matches the outer salt concentration. The shell has mostly planar anchoring on its surface, apparent from the defect required from concentric cholesteric layers (zoomed on right).}
\end{figure*}

\begin{figure*}[ht!]
\centerline{\includegraphics[width=0.5\textwidth]{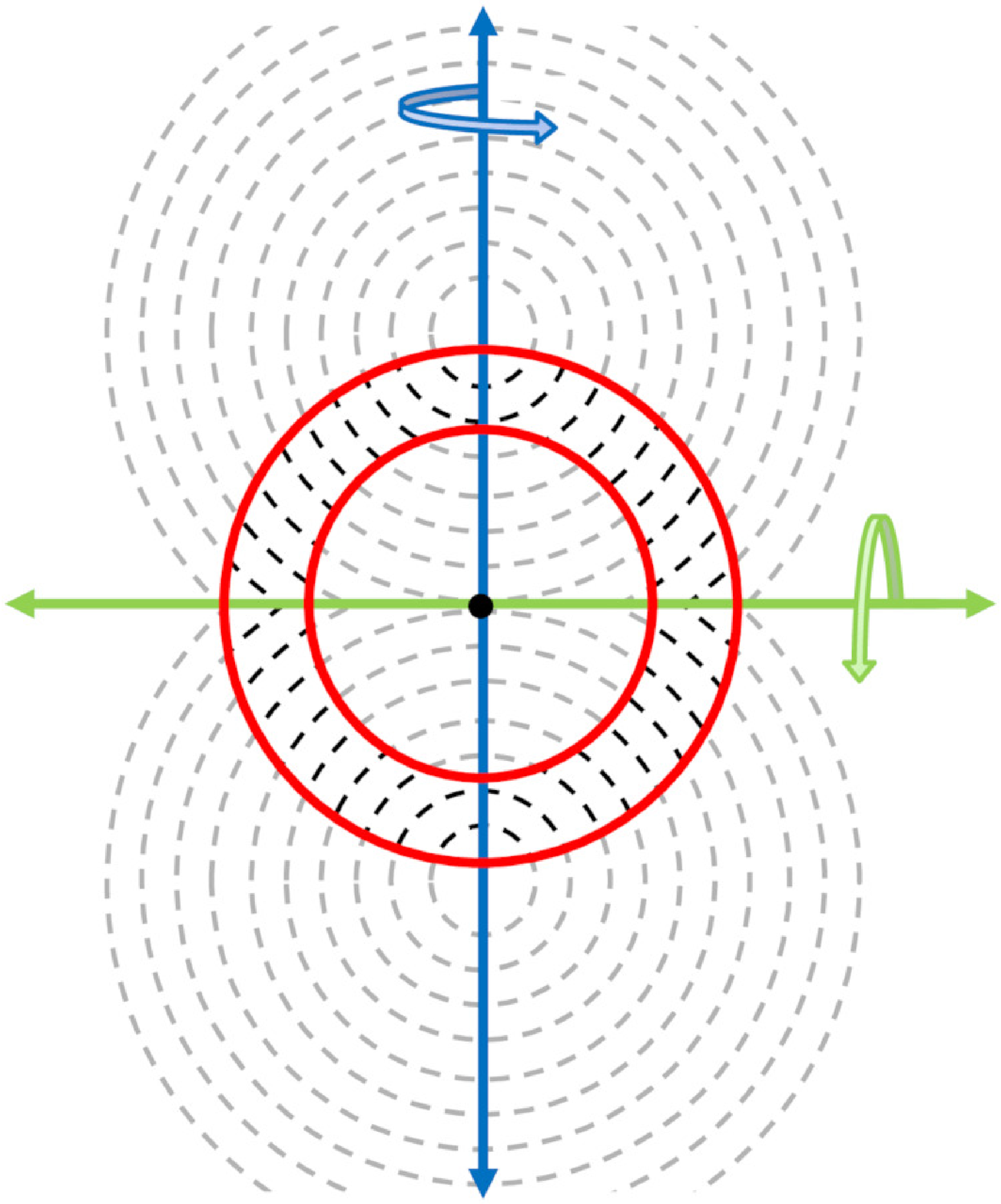}}
\caption{\label{FCDDiagram} \textbf{Double spirals are signatures of focal conic domains (FCDs).} Layers of concentric circles (dashed lines), representative of cholesteric layers, collide on a focal line (green). In red, we depict the surfaces of the shell confining the cholesteric, separating physical areas of the circular sections (black dashed lines) from virtual areas (gray dashed lines). If we rotate the two-dimensional texture around the vertical axis (blue) we get a texture between two red spheres with a focal plane dividing the top and bottom concentric CLC spheres. Topology, however, forces in two charge +1 nematic defects along the rotation axis. Though a +1 disclination in a uniaxial nematic can ``escape into the third dimension'' \cite{cladiskleman}, the biaxial-like cholesteric texture cannot \cite{geomchol}. As a result, the nematic disclinations become pitch defects, located along the rotation axis. Alternatively, were we to rotate the two-dimensional texture around the horizontal axis (green), we generate a toroidal focal conic \cite{pieranski}, with one focal line connecting the centers of the concentric CLC circles and another focal line along the axis of rotation. Pitch defects, at the center of double spirals, are located along the green axis of rotation. The director configuration within the shell is similar for either of these rotation axes. The structure of the FCDs becomes more complex as the number of double spiral domains increases.}
\end{figure*}

\begin{figure*}[ht!]
\centerline{\includegraphics[width=0.9\textwidth]{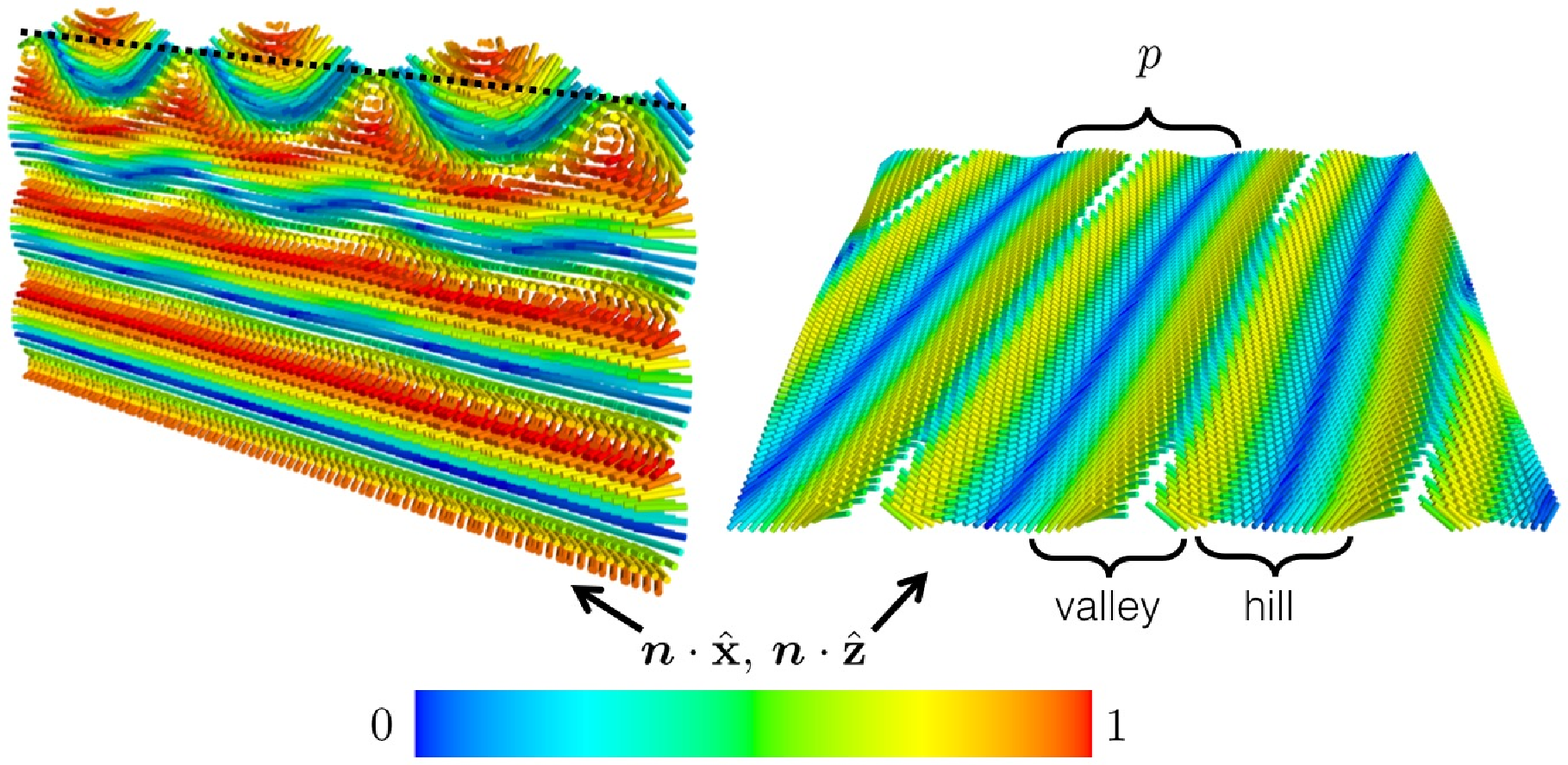}}
\caption{\label{Coexistence} \textbf{Interface undulations at an isotropic-CLC interface.}  We simulate an isotropic-CLC interface in a slab geometry ($24 \times 24$ nm in the $x$ and $y$ directions and $18$ nm wide in the vertical direction) by tuning our Landau-de Gennes potential to the coexistence point, which corresponds to constants $A= 5.2311$, $B=-11.2612$, and $C=0.8889$.  In this case, the CLC and isotropic phases are equally favored energetically.   We plot just the CLC\ region, showing the local orientation of the molecules according to the $\bm{Q}$ tensor. By starting with an initial condition that is isotropic on the top of the slab and a uniform CLC on the bottom, we make an interface where the two regions initially meet in the middle of the slab. The anchoring conditions are strongly planar on the bottom of the slab (to force the pitch to align along the $z$-axis in the bulk of the slab) and free on the top. There are periodic boundary conditions in the $x$ and $y$ directions.  After minimizing the Landau-de Gennes free energy, we see that the isotropic-CLC interface undulates (with hills and valleys indicated) and forms stripes with a periodicity equal to one pitch.  This illustrates that a free CLC interface naturally undulates in response to the variation of the molecular orientation at the interface.  We use the two-constant approximation with $L_1=L_2=2.32$ and $L_{24}=L_2/2$ (see SI, Materials and Methods).  The pitch is 6 nm in the CLC region. }
\end{figure*}

\begin{figure*}
\centering
  \includegraphics[width=1\textwidth]{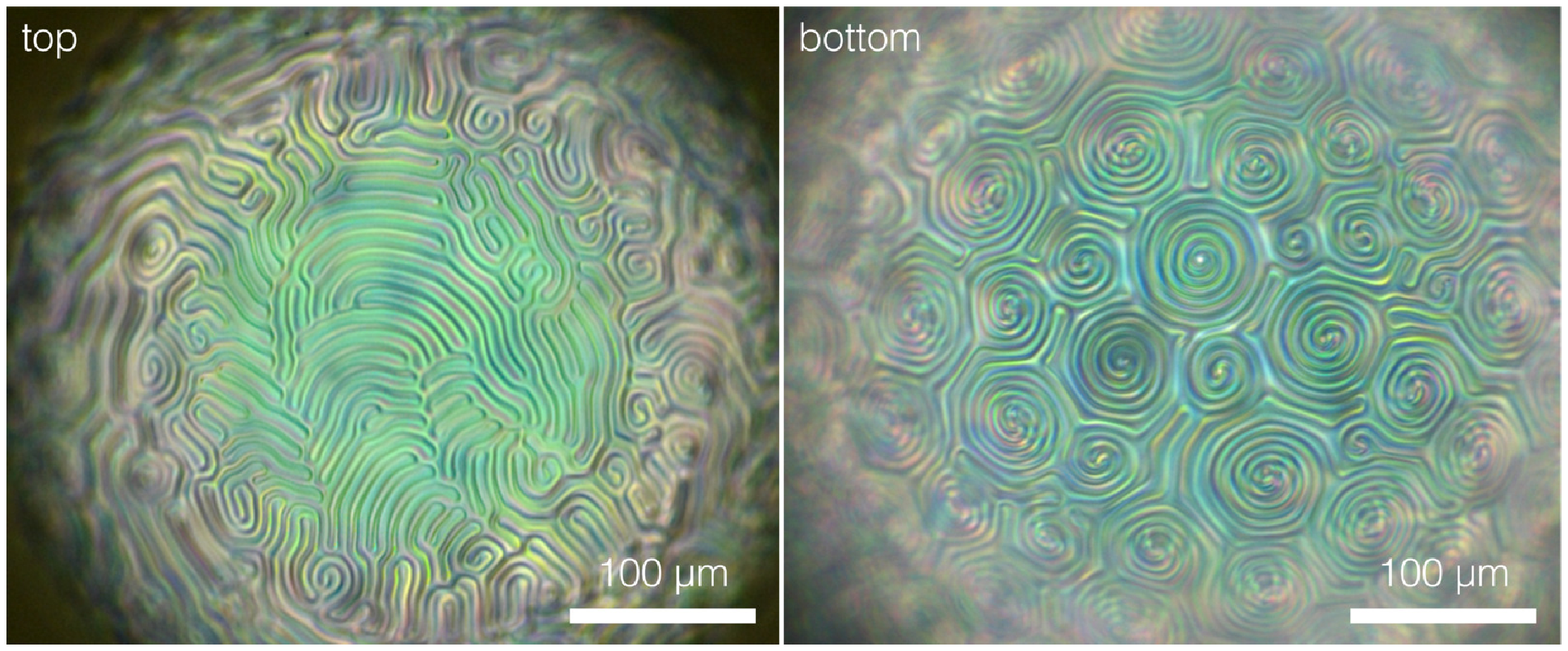}
  \caption{\label{BentStateFCD} \textbf{``Bent'' state can coexist with FCD state in CLC shell with heterogeneous thickness.} A CLC shell with 0 mM SDS, 1 M NaCl, and 1\% PVA in the inner water phase begins to swell in an outer water solution with 7 mM SDS, 0.1 M NaCl, and 1\% PVA. Thickness variation in the CLC shell shows that thinner regions have stripes from bent cholesteric layers (left), expressing the ``bent'' state, while thicker regions in the back of the drop have FCDs (right).}
\end{figure*}

\indent \indent \textbf{SI Video 1: Surfactant anchoring transition - strong homeotropic to planar.} A cholesteric liquid crystal shell is first equilibrated with no SDS in the inner water phase and 10 mM SDS in the outer phase. Both water phases have 1\% PVA and 0.1 M NaCl. The shell is then transferred to a solution with no SDS, but still with 1\% PVA and 0.1 M NaCl. As the SDS diffuses from the outer surface of the shell and into the outer water phase, the anchoring on the shell weakens, causing the shell patterns to change. Subsurface defect lines become focal conic domains, then thick planar stripes, until the planar anchoring state is reached. The pitch is 5 $\mu$m. 

\textbf{SI Video 2: Temperature anchoring transition - 3\% CB15, thick shell.} A cholesteric liquid crystal shell with only 1\% PVA in the water phases under goes a temperature ramp towards the cholesteric-isotropic transition temperature at a rate of 0.01 $^{\circ}$C/minute. The absolute value of temperature is given on every frame of the video. The pitch is 5 $\mu$m. Wide, planar stripes form with the ``bent''-state, then the stripe width halves discontinuously to the thin stripe state, until the shell fully transitions to isotropic. The shell is 32 $\mu$m thick, $c>1$ initially. Since many pitches can fit into the shell thickness, the cholesteric layers can easily bend to form focal conic domains. Hexagonal packing of the FCDs occur first, then the polygonal texture forms, until the shell thins, becoming a 2D nematic before fully transitioning to the isotropic phase. 

\textbf{SI Video 3: Temperature anchoring transition - 3\% CB15, thin shell.} A cholesteric liquid crystal shell with only 1\% PVA in the water phases undergoes a temperature ramp towards the cholesteric-isotropic transition temperature at a rate of 0.01 $^{\circ}$C/minute. The relative value of temperature to the phase transition temperature is given on every frame of the video. The pitch is 5 $\mu$m. The shell thickness is comparable to or less than the pitch, $c<1$. Wide, planar stripes form with the ``bent''-state, then the stripe width halves discontinuously to the thin stripe state. The shell is then so thin that it is essentially a 2D nematic, until the shell fully transitions to isotropic. The accompanying graph shows how the stripe periodicity changes with time from measuring intensity differences. The discontinuity in the periodicity when the CLC becomes very thin is apparent in the center plot of periodicity/pitch vs. $\Delta T(^{\circ} C)$ from the clearing point.

\textbf{SI Video 4: Equilibrium to planar transition, 1.} Video of Fig.~\ref{EqThinStripes}C. The stripe dislocation becomes the planar defect.

\textbf{SI Video 5: Equilibrium to planar transition, 2.} SDS is removed from the outer cholesteric shell surface, decreasing the homeotropic anchoring strength and unwinding the stripes. The stripe patterns match those of Fig.~\ref{EqThinStripes}C and SI Video 4. The stripe dislocation becomes the planar defect. 

\textbf{SI Video 6: Equilibrium to planar transition, 3.} As SDS diffuses from the outer cholesteric shell surface to the surrounding water phase, the homeotropic anchoring strength decreases and the stripes unwind. The top of the shell is thinner than in the previous two surfactant anchoring transitions from equilibrium thin stripes, evident from the Skyrmions at the top of the shell. As the stripes unwind, a Skyrmion becomes a planar defect. A Skyrmion is a pitch defect, just as a stripe dislocation is a pitch defect, making it an energetically favorable place for the planar defect to form.

% \bibliography{StripedSphereBib}
\bibliographystyle{unsrt}
%\printbibliography

\clearpage

\end{document}